\documentclass[nofootinbib, superscriptaddress]{revtex4-2}
\usepackage[utf8]{inputenc}
\usepackage{bm, amsthm, amsmath, amsfonts, amssymb, color, graphicx, natbib}
\usepackage{enumitem}
\usepackage{slashed}
\usepackage{xcolor}
\usepackage{verbatim}
\usepackage{float}
\graphicspath{{fig/}}
\usepackage{hyperref}
\hypersetup{
	colorlinks=true,
	linkcolor=blue,
	filecolor=blue,      
	urlcolor=blue,
	citecolor=blue
}

\begin{document}
	
\title{The merger rate of primordial black hole binaries as a probe of Hubble parameter} 
	
\author{Qianhang Ding}
\email{dingqh@ibs.re.kr}
	
\affiliation{Cosmology, Gravity and Astroparticle Physics Group, Center for Theoretical Physics of the Universe,\\
	Institute for Basic Science, Daejeon 34126, Republic of Korea}
	
\begin{abstract}
	We propose that the merger rate of primordial black hole (PBH) binaries can be a probe of Hubble parameter by constraining PBH mass function in the redshifted mass distribution of PBH binaries. In next-generation gravitational wave (GW) detectors, the GWs from PBH binaries would be detected at high redshifts, which gives their redshifted mass and luminosity distances. From a number of detected events, the redshifted mass distribution of PBH binaries can be statistically obtained, and it depends on PBH mass function and redshift distribution of detected PBH binaries. The PBH mass function can be inversely solved by applying the gradient descent method in the relation between redshifted mass distribution and redshift distribution. However, the construction of redshift distribution requires an assumed Hubble parameter in a background cosmology to extract redshift from luminosity distances, which causes solved PBH mass function also depends on assumed Hubble parameter. To determine the Hubble parameter, the merger rate of PBH binaries constrains on this Hubble parameter-dependent PBH mass function by comparing calculated merger rate distribution with observed one, and the best-fit result produces an approximate mass distribution of the physical PBH mass function and pins down the Hubble parameter. 
\end{abstract}
	
\maketitle

\section{Introduction}

The value of Hubble parameter is of central importance in modern cosmology. However, cosmological observations report an inconsistency in the measurement value of Hubble parameter. The local measurement of Hubble parameter gives a local value of $H_0 = 73.04 \pm 1.04 \, \mathrm{km} \, \mathrm{s}^{-1} \, \mathrm{Mpc}^{-1}$ \cite{Riess:2021jrx} (also see \cite{Riess:2023egm} for a review), which is $5 \sigma$ larger than the measurement value of Hubble parameter $H_0 = 67.4 \pm 0.5  \, \mathrm{km} \, \mathrm{s}^{-1} \, \mathrm{Mpc}^{-1}$ from Planck 2018 result under Lambda-cold dark matter ($\Lambda$CDM) model \cite{Planck:2018vyg}. This is so-called Hubble tension and its $5 \sigma$ inconsistency has become a cosmological crisis in $\Lambda$CDM cosmology \cite{Perivolaropoulos:2021jda, Abdalla:2022yfr, DiValentino:2021izs}. 

To reveal Hubble tension and toward a concordance cosmological model, the dynamical evolution of Hubble parameter from the early universe to the late universe should be well measured. Currently, for early universe probes, cosmic microwave background (CMB) provides a Hubble parameter measurement at redshift $z \simeq 1100$ by fitting CMB power spectrum \cite{Planck:2018vyg}. For late universe probes, Type Ia supernovae (SNe) provide a Hubble parameter measurement at redshift $z < \mathcal{O}(1)$ by constraining redshift-luminosity distance relation \cite{Fernie_1969, Pan-STARRS1:2017jku}, which works as standard candles. And baryon acoustic oscillation (BAO) can measure Hubble parameter at redshift $z \sim \mathcal{O}(1)$ by calibrating redshift-angular diameter distance relation  \cite{beutler20116df, Ross:2014qpa, BOSS:2014hwf}, which is so-called standard rulers. In the intermediate epoch between early universe and late universe, gravitational waves (GWs) from compact binaries and their electromagnetic counterparts can be a potential Hubble parameter probe up to redshift $z \sim \mathcal{O}(10)$ as a standard siren in next-generation GW detectors \cite{Schutz:1986gp, Holz:2005df, LIGOScientific:2017vwq, LIGOScientific:2017adf}. However, Hubble parameter measurement between redshift $z \sim \mathcal{O}(10)$ and $z \sim \mathcal{O}(1000)$ is still not promising, due to lack of enough cosmological signals in the dark age \cite{Miralda-Escude:2003lmt}, which causes difficulty in constructing redshift-distance relation to extract Hubble parameter.

The possible existence of primordial black hole (PBH) binaries changes the story. PBHs are hypothetical astrophysical objects, which were born from primordial perturbations in highly overdense regions by gravitational collapse \cite{Hawking:1971ei, Zeldovich:1967lct, Carr:1974nx, Khlopov:2008qy}. The formation of PBH binaries occurs in early universe epoch, like radiation-domination era or matter-radiation equality, depends on PBH mass and abundance \cite{Sasaki:2016jop, Ali-Haimoud:2017rtz}. The GWs emitted from PBH binaries could propagate cosmological distance from the early epoch to the present, where luminosity distance is encoded in their GW waveform. Although redshift information is still missing, due to no electromagnetic counterpart of PBH binaries in the dark age \cite{Miralda-Escude:2003lmt}, and causes difficulty in breaking the mass-redshift degeneracy in GW waveform \cite{Schutz:1986gp}. The unique statistic properties of PBH binaries open another window in cosmic probes, the statistic distribution of PBH binaries, such as PBH mass function and probability distribution on orbital parameters, can provide an additional constraint on degenerate parameters to help extract their intrinsic values, for instance, studying the evolution of probability distribution on semi-major axis and eccentricity of PBH binaries from the initial state to a later state can construct a redshift-time relation, which constrains the Hubble parameter, this is one kind of standard timers \cite{Ding:2022rpd, Cai:2021fgm}. Due to the possible existence of PBH binaries in cosmic history, their unique statistic properties can help PBH binaries become a potential probe in tracking cosmic evolution.

In this paper, we focus on using the merger rate of PBH binaries to put a constraint on PBH mass function, which leads to a measurement of Hubble parameter. From a number of detected PBH binaries at different redshifts, the redshifted mass of PBHs and luminosity distance can be extracted from the GW waveform of each PBH binary. We can statistically construct a redshifted mass distribution of PBH binaries based on extracted redshifted PBH mass. Meanwhile, cosmic redshifts can be solved from obtained luminosity distances under an assumed Hubble parameter in a given cosmology such as $\Lambda$CDM model. Then a redshift distribution of detected PBH binaries is statistically constructed from solved redshifts. Combine the redshifted PBH mass distribution and redshift distribution, a Hubble parameter-dependent PBH mass function can be reconstructed. To determine the Hubble parameter, the merger rate of PBH binaries put a constraint on PBH mass function \cite{Chen:2018czv}, we can calculate a merger rate distribution of PBH binaries based on derived Hubble parameter-dependent PBH mass function and compared it with observed one. The fitting result of merger rate distribution of PBH binaries depends on this Hubble parameter-dependent PBH mass function, when assumed Hubble parameter is approaching its physical value, solved PBH mass function is approaching its physical mass distribution, this comparison can approach the best-fit result, which breaks the degeneracy between PBH mass function and redshift distribution, and pins down the correct Hubble parameter. 

Although, we have not observed any PBH binaries so far, there still existing some GW events which may hint the existence of PBH binaries, e.g., GW190425 and GW190814 show a companion mass of compact binary is smaller than $3 \, M_\odot$ \cite{LIGOScientific:2020aai, LIGOScientific:2020zkf}, which could be candidates of PBHs \cite{Clesse:2020ghq}, GW190521 reports a compact binary event with mass in astrophysical BH mass gap \cite{LIGOScientific:2020iuh}, which can also be explained in PBH scenarios \cite{Clesse:2020ghq, DeLuca:2020sae}. In the next-generation GW detectors, such as Laser Interferometer Space Antenna (LISA) \cite{bender1998lisa} and Big Bang Observer (BBO) \cite{Harry:2006fi}, can detect GWs at redshift $z > 20$, where astrophysical BHs have not formed, therefore the detection of high redshift compact binaries could be a smoking gun for the existence of PBH binaries \cite{Nakamura:2016hna, Ding:2020ykt}.

This paper is organized as follows. In Sec.~\ref{sec:redshifted_mass_dis}, we construct the connection among redshifted mass distribution of PBH binaries, PBH mass function and redshift distribution of detected PBH binaries. In Sec.~\ref{sec:inverse}, we apply the gradient descent method in redshifted mass distribution of PBH binaries to inversely solve PBH mass function. In Sec.~\ref{sec:Hubble_depend}, we calculate the dependence of redshift distribution of detected PBH binaries on Hubble parameter, which leads to a Hubble parameter-dependent PBH mass function in the inverse process. In Sec.~\ref{sec:merger_rate}, we use the merger rate distribution of PBH binaries to constrain the Hubble parameter-dependent PBH mass function, which works as a Hubble parameter probe. In Sec.~\ref{sec:conclusion_discussion}, the conclusion and discussions on this Hubble parameter probe are given.

\section{Redshifted mass distribution of PBH binaries}\label{sec:redshifted_mass_dis}

In next-generation GW detectors, the detectable redshift range of BH binaries would be larger than $20$, which provides an observational window for PBH binaries  \cite{Ding:2020ykt}. From GW waveform of PBH binaries, key properties like redshifted chirp mass $\mathcal{M}_z$, mass ratio $q$, luminosity distance $d_L$, etc., can be extracted. Then an observed redshifted mass distribution of PBH binaries $P(m_1^{z}, m_2^{z})$ can be statistically constructed, which is defined as follows,
\begin{align}
	P(m_1^z, m_2^z) = \frac{1}{N_\mathrm{tot}} \frac{dN_\mathrm{obs}(m_1^z, m_2^z)}{dm_1^z \, dm_2^z}~,
\end{align}
where $dN_\mathrm{obs}(m_1^z, m_2^z)$ is the number of observed PBH binaries within redshifted mass range $(m_1^z, m_1^z + dm_1^z)$ and $(m_2^z, m_2^z + dm_2^z)$, $N_\mathrm{tot}$ is the total number of PBH binaries, $m_1^z$ and $m_2^z$ are the redshifted mass of each PBH component in binaries, that can be found via observable quantities as follows,
\begin{align}
	m_1^z = \mathcal{M}_z q^{2/5} (1+q)^{1/5} ~,~~~m_2^z = \mathcal{M}_z \frac{(1+q)^{1/5}}{q^{3/5}}~.
\end{align}
Intuitively, the observed redshifted mass distribution of PBH binaries depends on PBH mass function, sensitivity of GW detectors, and redshift distribution of detected PBH binaries. To find this relation, we start from calculating the cumulative distribution of redshifted PBH mass $C(m_1^z, m_2^z)$, which is defined as follows,
\begin{align}
	C(m_1^z, m_2^z) = \frac{N_\mathrm{obs}(m_a < m_1^z, m_b < m_2^z)}{N_\mathrm{tot}}~,
\end{align}
where $N_\mathrm{obs}(m_a < m_1^z, m_b < m_2^z)$ is the number of observed PBH binaries with the redshifted mass of their components smaller than $m_1^z$ and $m_2^z$. And it can be expressed in following form (for detailed derivation, see Appendix.~\ref{app:PBH_dis}),
\begin{align}\label{eq:cumulative}
	C(m_1^z, m_2^z) &= \int_0^{m_1^z} \int_0^{m_2^z} P(m_a, m_b) \, d m_a \, d m_b \nonumber\\
	&=\int_0^\infty \int_0^{\frac{m_1^z}{1+z}} \int_0^{\frac{m_2^z}{1+z}} n(m_1) n(m_2)  W(m_1, m_2; z) p(z) \, dm_1 \, dm_2 \, dz~.
\end{align}
Here, $n(m)$ is the PBH mass function, which is defined such that the quantity $n(m) dm$ represents the probability that a randomly chosen PBH in the mass range $(m, m+dm)$, and it can be expressed in form of PBH number density as follows,
\begin{align}
	n(m) = \frac{1}{n_\mathrm{PBH}} \frac{dn}{dm}~,
\end{align}
where $dn$ is comoving number density of PBHs in the mass range $(m, m+dm)$ and $n_\mathrm{PBH} = \int dn$ is the total comoving number density of PBHs. This PBH mass function $n(m)$ is normalized as $\int n(m) dm = 1$. $W(m_1, m_2; z)$ is the detectable window function of PBH binaries with intrinsic mass $m_1$ and $m_2$ at redshift $z$, and $p(z)$ is the redshift distribution of detected PBH binaries. From Eq.~\eqref{eq:cumulative}, we can obtain the redshifted mass distribution of PBH binaries by derivating $C(m_1^z, m_2^z)$ with respect to redshifted mass $m_1^z$ and $m_2^z$ as follows (see Eq.~\eqref{appeq:cumu_1} and \eqref{appeq:cumu_2} for details),
\begin{align}\label{eq:probability}
	P(m_1^z, m_2^z) &= \frac{\partial^2 C(m_1^z, m_2^z)}{\partial m_1^z \, \partial m_2^z}\nonumber\\ 
	&= \int_0^\infty n(\frac{m_1^z}{1+z}) n(\frac{m_2^z}{1+z}) W(\frac{m_1^z}{1+z}, \frac{m_2^z}{1+z}; z) \frac{p(z)}{(1+z)^2} \, dz~.
\end{align}
In order to connect redshifted mass distribution of PBH binaries $P(m_1^z, m_2^z)$ with PBH mass function $n(m)$, we need to figure out the detectable window function $W(m_1, m_2; z)$ and redshift distribution $p(z)$. Then the relation between $P(m_1^z, m_2^z)$ and $n(m)$ can be obtained and further can be applied in an inverse process to solve $n(m)$ from $P(m_1^z, m_2^z)$.

Window function $W(m_1, m_2; z)$ denotes the detection probability of the observed PBH binaries among all existing PBH binaries with intrinsic mass $m_1$ and $m_2$ at redshift $z$, it comes from the selection effect due to limited sensitivity and frequency band of GW detectors, which causes only a part of PBH binaries with suitable orbital parameters, semi-major axis $a$ and eccentricity $e$, can be observed by given GW detectors \cite{Mandel:2018mve, Vitale:2020aaz, Gerosa:2020pgy}. The window function can be expressed as follows,
\begin{align}\label{eq:window_fun}
	W(m_1, m_2; z) = \frac{N_\mathrm{obs}(m_1, m_2; z)}{N_\mathrm{tot}(m_1, m_2; z)}~,
\end{align}
where $N_\mathrm{obs}(m_1, m_2; z)$ and $N_\mathrm{tot}(m_1, m_2; z)$ are the number of observed and total PBH binaries with intrinsic mass $m_1$ and $m_2$ at redshift $z$, respectively. The detection of a PBH binary requires a signal-to-noise ratio ($\mathrm{SNR}$) should exceed a threshold for a high confidence level, where we set the threshold value of $8$ \cite{Flanagan:1997sx}, which makes sure the detection probability is $> 95\%$, and a false alarm probability $< 0.1\%$. The SNR can be calculated as shown in \cite{Flanagan:1997sx, Rosado:2015voo},
\begin{align}\label{eq:SNR}
	\mathrm{SNR} = \sqrt{4 \int_{f_\mathrm{min}}^{f_\mathrm{max}} \frac{|\tilde{h}(f)|^2}{S_n(f)} df}~,
\end{align}
where $[f_\mathrm{min}, f_\mathrm{max}]$ is the observational frequency range. $S_n(f)$ is the noise strain of the detector, and $S_n(f)$ for each detector can be found from Fig.~A2 in \cite{Moore:2014lga}. In following parts, we adopt the $S_n(f)$ of BBO in calculations. $\tilde{h}(f)$ is the Fourier transformation of GW amplitude $h(t)$, which can be expressed in following form under the stationary approximation \cite{Droz:1999qx},
\begin{align}\label{eq:amplitude}
	\tilde{h}(f) = \sqrt{\frac{5}{24}} \frac{(G \mathcal{M}_c (1+z))^{5/6}}{\pi^{2/3} c^{3/2} d_L(z)} f^{-7/6}~,
\end{align}
where $G$ is the Newton's constant, $c$ is the speed of light, $\mathcal{M}_c = \mathcal{M}_z/1+z$ is the intrinsic chirp mass of PBH binaries, and $d_L$ is the luminosity distance to the PBH binary.

In order to evaluate Eq.~\eqref{eq:window_fun}, we need to know the detectable orbital parameters of the PBH binaries with intrinsic mass $m_1$ and $m_2$ at redshift $z$, and then integrate the PBH binaries probability distribution on semi-major axis and eccentricity $P(a, e; z)$ over detectable parameter region as follows,
\begin{align}\label{eq:windowfun_cal}
	W(m_1, m_2; z) = \int_{a_\mathrm{min}}^{a_\mathrm{max}} \int_{e_\mathrm{min}}^{e_\mathrm{max}} P(a, e; z) da de =  \int_{a_\mathrm{min}^\mathrm{ini}}^{a_\mathrm{max}^\mathrm{ini}} \int_{e_\mathrm{min}^\mathrm{ini}}^{e_\mathrm{max}^\mathrm{ini}} P(a, e; z_\mathrm{ini}) da de~,
\end{align}
where $a_\mathrm{min}$ ($e_\mathrm{min}$) and $a_\mathrm{max}$ ($e_\mathrm{max}$) are minimal and maximal detectable orbit parameters, respectively. Their values can be numerically solved from the detection requirement $\mathrm{SNR} > 8$ in Eqs.~\eqref{eq:SNR} and \eqref{eq:amplitude}. We mainly focus on the detection of small eccentricity orbits ($e \sim 0$) due to lack of suitable high-eccentricity GW waveform template \cite{Gayathri:2020coq}. The superscript $\mathrm{ini}$ on $a$ and $e$ denotes the initial value of semi-major axis and eccentricity when PBH binaries form.  The initial value of $a$ and $e$ can be solved by the formula in \cite{Peters:1964zz} as follows,
\begin{align}\label{eq:peter}
	\frac{d a}{d t} = - \frac{64}{5} \frac{G^3 m_1 m_2 (m_1 + m_2)}{c^5 a^3 (1 - e^2)^{7/2}} \left(1 + \frac{73}{24} e^2 + \frac{37}{96} e^4\right)~,~~~
	\frac{d e}{d t} = - \frac{304}{15} \frac{G^3 m_1 m_2 (m_1+m_2)}{c^5 a^4 (1-e^2)^{5/2}} e \left(1+\frac{121}{304} e^2\right)~.
\end{align}
Then we solve Eq.~\eqref{eq:peter} backward in time from the observable parameters $(a, e)$ at redshift $z$ to the initial parameters $(a^\mathrm{ini}, e^\mathrm{ini})$ at PBH binary formation redshift $z_\mathrm{ini}$ in Eq.~\eqref{appen_eq:formation_redshift}. The probability distribution on semi-major axis and eccentricity at initial redshift $z_\mathrm{ini}$ for PBH binaries with intrinsic mass $m_1$ and $m_2$ can be approximated as \cite{Sasaki:2016jop, Ali-Haimoud:2017rtz}
\begin{align}\label{eq:initial_distribution}
	P(a, e, z_\mathrm{ini}) = \frac{3}{4} f_b^{3/2} \frac{a^{1/2}}{\bar{x}^{3/2}} \frac{e}{(1-e^2)^{3/2}}~,
\end{align}
where $f_b$ is the energy density fraction of PBH binaries with mass $m_1$ and $m_2$ in the dark matter and $\bar{x}$ is the physical mean separation of PBH binaries with intrinsic mass $m_1$ and $m_2$ at matter-radiation equality, see Eq.~\eqref{eq:mean_distance} and \eqref{appen_eq:mean_dist}. Combine Eqs.~(\ref{eq:window_fun} -- \ref{eq:initial_distribution}), window function $W(m_1, m_2; z)$ for PBH binaries with intrinsic mass $m_1$ and $m_2$ at redshift $z$ can be numerically calculated.

Redshift distribution $p(z)$ describes the distribution of detected PBH binaries at different redshifts, which is defined as follows,
\begin{align}
	p(z) = \frac{1}{N_\mathrm{obs}}\frac{dN_\mathrm{obs}(z)}{dz}~,
\end{align}
where $dN_\mathrm{obs}(z)$ is the number of observed PBH binaries within redshift range $(z, z+dz)$. $p(z)$ is normalized as $\int p(z) dz = 1$, and $N_\mathrm{obs}$ denotes the total number of observed PBH binaries. It depends on detectable event rate of PBH binaries, comoving volume, and selection effect in GW detectors, at different redshifts. In order to obtain this redshift distribution $p(z)$, we can statistically count detected PBH binary events at different redshifts in observations. Theoretically, we use normalized redshift distribution of PBH merger rate population as an approximation as shown in \cite{Ng:2022agi},
\begin{align}\label{eq:redshift_dis}
	p(z) \propto \frac{\dot{n}(z)}{1+z} \frac{dV_c}{dz}~,
\end{align}
where we set a detectable redshift range $20 \leq z \leq 100$ as an approximation of the selection effect on redshift. $\dot{n}(z)$ is the volumetric merger rate density of PBHs, which is power-law dependent on the age of the universe $t(z)$ as $\dot{n}(z) \propto (t(z)/t_0)^{-34/37}$ \cite{Raidal:2018bbj}, where $t_0$ is the current age of the universe. $1+z$ term comes from the cosmic time dilation. $dV_c/dz$ is the differential comoving volume, which can be calculated as $dV_c/dz = 4 \pi D_H D_M^2/E_z(z) $ in flat universe, where $D_H = c/H_0$ is Hubble distance, $D_M$ is transverse comoving distance, and $E_z(z) = \sqrt{\Omega_\mathrm{M}(1+z)^3 +\Omega_\Lambda}$ in flat $\Lambda$CDM model \cite{Hogg:1999ad}.

\begin{figure}[htbp] \centering
	\includegraphics[width=8cm]{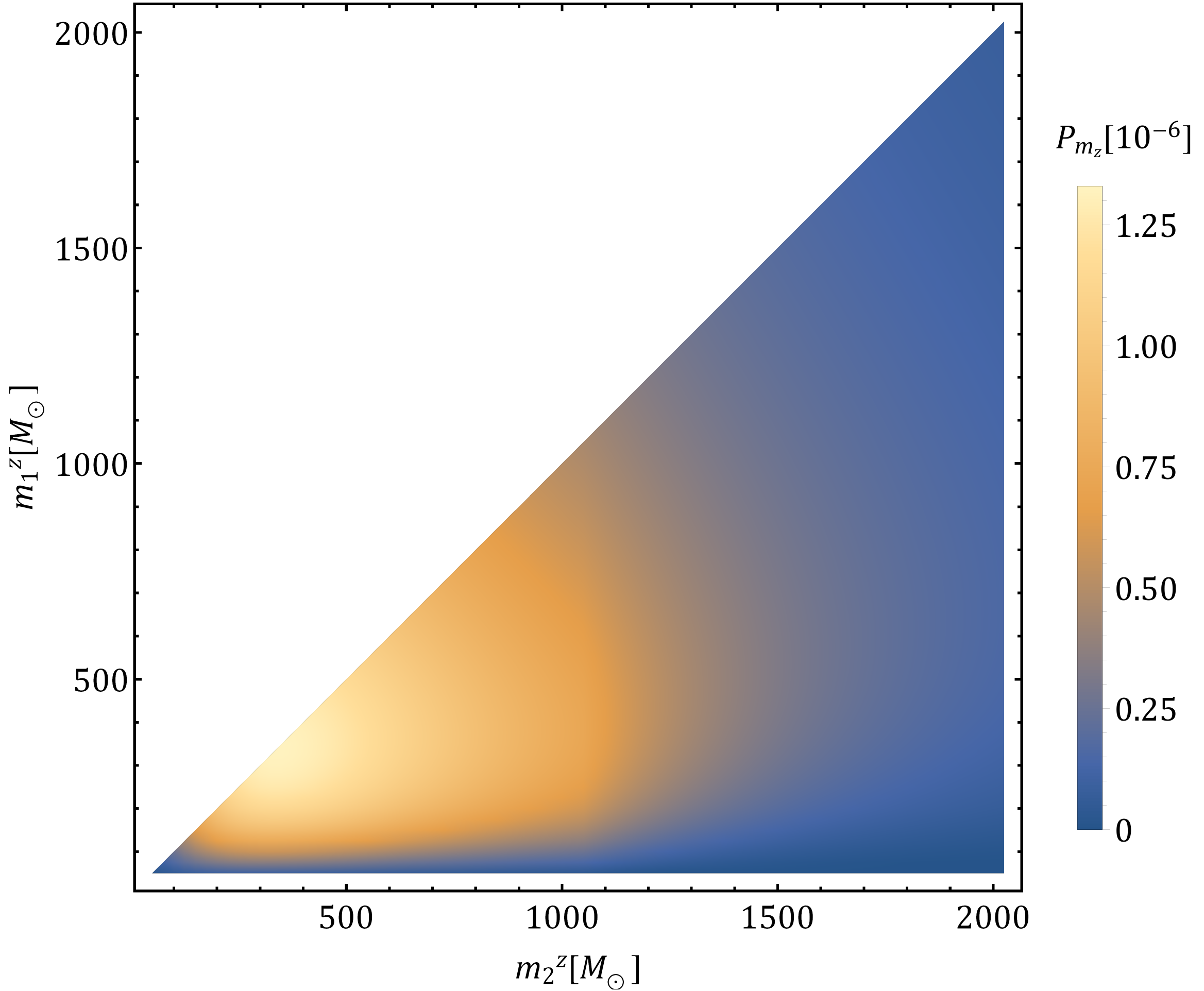}
	\includegraphics[width=8cm]{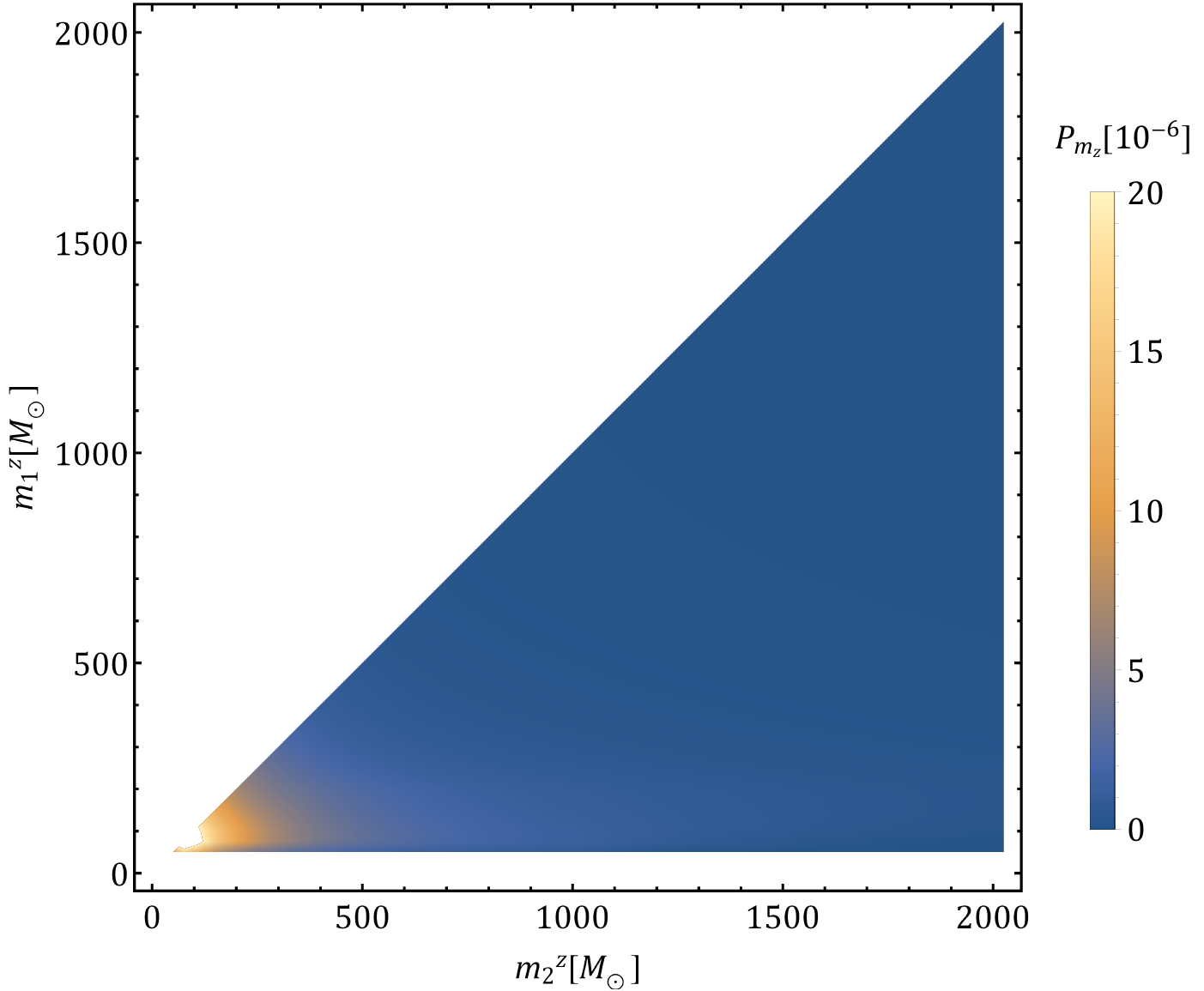}
	\caption{\label{fig:redshift_distribution}
		\textit{Left panel}: The normalized redshifted mass distribution of PBH binaries for a log-normal PBH mass function with $m_c = 30 \, M_\odot$ and $\sigma = 1$. \textit{Right panel}: The normalized redshifted mass distribution of PBH binaries for a power-law PBH mass function with $\alpha = 1.5$ and $M = 2 M_\odot$. The assumed cosmology for both panels is a flat $\Lambda$CDM model with Planck 2018 parameters $\Omega_\mathrm{M} = 0.315$, $\Omega_\Lambda = 0.685$ and $H_0 = 67.4 \, \mathrm{km} \, \mathrm{s}^{-1} \, \mathrm{Mpc}^{-1}$ \cite{Planck:2018vyg}, and the redshifted mass $m_1^z$ and $m_2^z$ is set to follow $m_1^z \leq m_2^z$. 
	}
\end{figure}

Taking account Eqs.~\eqref{eq:windowfun_cal} and \eqref{eq:redshift_dis} in Eq.~\eqref{eq:probability}, and we consider two typical PBH mass functions, log-normal distribution \cite{Dolgov:1992pu} and power-law distribution \cite{Carr:1975qj}
\begin{align}
	n_\mathrm{LN}(m) = \frac{1}{\sqrt{2 \pi} \sigma m} \exp \left(- \frac{\log^2(m/m_c)}{2 \sigma^2}\right)~,~~~n_\mathrm{PL}(m) = \frac{\alpha - 1}{M}\left(\frac{m}{M}\right)^{-\alpha}~,
\end{align}
where $\sigma$ denotes the mass width of log-normal distribution, $m_c$ is the critical mass in log-normal distribution, and $\alpha$ is the power index in power-law distribution and $M$ is the minimal mass in power-law distribution. Then we can obtain the redshifted mass distribution of PBH binaries as shown in Fig.~\ref{fig:redshift_distribution}.
It shows the redshifted mass distribution of PBH binaries strongly depends on PBH mass functions, and we can find the redshifted mass distribution with a log-normal PBH mass function spreads broadly, while a power-law PBH mass function produces a narrow redshifted mass distribution that is centered around a small value of redshifted mass. This indicates the redshifted mass distribution of PBH binaries could be a potential tool in distinguishing different PBH mass functions.

In above discussion, we mainly focus on the preliminary relation between the redshifted mass distribution $P(m_1^z, m_2^z)$ and PBH mass function $n(m)$. However, late time evolution of PBHs would affect observed redshifted mass distribution, such as PBH mass increment due to matter accretion from surrounding environments \cite{DeLuca:2020fpg}, PBH binaries disruption by matter inhomogeneities \cite{Raidal:2018bbj}, and PBH binaries formation by direct capture and three-body interaction inside the clusters \cite{Franciolini:2022ewd}. These effects would modify the PBH mass function at late time and observed number density of PBH binaries, so we need to take all of them into account to construct a relation between the observed redshifted mass distribution and primordial PBH mass function (see Appendix.~\ref{app:modification} for details), which we leave it in future studies. In following sections, we use above preliminary relation to show how this proposed method probe the Hubble parameter.

\section{PBH mass function from inverse redshifted mass distribution}\label{sec:inverse}

Theoretically, we can calculate the redshifted mass distribution of PBH binaries based on PBH mass function $n(m)$, detector-dependent window function $W(m_1, m_2; z)$ and the redshift distribution of detected PBH binaries $p(z)$. Observationally, we can only detect the redshifted mass distribution of PBH binaries. The relation between redshifted mass distribution of PBH binaries and PBH mass function follows Eq.~\eqref{eq:probability} as
\begin{align}\label{eq:probability_1}
	P_O(m_1^z, m_2^z) = \int_0^\infty n_p(\frac{m_1^z}{1+z}) n_p(\frac{m_2^z}{1+z}) W(\frac{m_1^z}{1+z}, \frac{m_2^z}{1+z}; z) \frac{p(z)}{(1+z)^2} \, dz~,
\end{align}
where $P_O(m_1^z, m_2^z)$ with subscript $O$ denotes the observed redshifted mass distribution and $n_p(m)$ is the physical PBH mass function. Apart from $n_p(m)$, $P_O(m_1^z, m_2^z)$ only depends on detector-dependent window function $W(m_1, m_2; z)$ and redshift distribution of detected PBH binaries $p(z)$. For a given GW detector, the selection effect in window function is fixed. Also, redshift distribution can be constructed based on the observable from observed PBH binaries, where GW waveform of PBH binaries provides the luminosity distances between PBH binaries and the earth, if we know the physical cosmological model with a fixed Hubble parameter, redshift can be solved from obtained luminosity distance, then redshift distribution of detected PBH binaries can be statistically constructed. After knowing the contribution of $W(m_1, m_2; z)$ and $p(z)$ in Eq.~\eqref{eq:probability_1}, the physical PBH mass function $n_p(m)$ can be inversely solved from observed redshifted mass distribution $P_O(m_1^z, m_2^z)$.

To obtain PBH mass function from inverse redshifted mass distribution, we apply the gradient descent method (see \cite{ruder2016overview} for detailed introductions) in Eq.~\eqref{eq:probability_1}. There are several steps in the gradient descent method as follows,
\begin{itemize}[itemsep=1pt,topsep=5pt,parsep=1pt]
	\item[1.] Give an initial PBH mass distribution as an initial condition of PBH mass function.
	\item[2.] Calculate theoretical redshifted mass distribution and compare it with observed one to obtain the error function.
	\item[3.] Update PBH mass function by calculating the gradient of error function with respect to the variation of PBH mass function.
	\item[4.] Iterate PBH mass function until finding the minimal value of error function, which is a best-fit approximation of physical PBH mass function.
\end{itemize}

In order to decrease time complexity of this algorithm, we start from discretizing PBH mass function $n(m)$ to a PBH mass distribution vector $\mathbf{n}$, where the component of vector follows $\mathbf{n}_\mathrm{i} = n(m_i)$ and a set of $\{m_i| 1 \leq i \leq N\}$ is $N$ mass points taken from a suitable PBH mass range. Then we provide an initial value of this PBH mass distribution vector $\mathbf{n}$. This initial vector could follow arbitrary mass distribution, however, choosing a more realistic and reasonable initial mass distribution vector could effectively decrease the time complexity of the gradient descent method. After giving the initial value of $\mathbf{n}$, an initial PBH mass function can be approximately constructed by interpolating the initial PBH mass distribution vector.

For a given assumed PBH mass function, we can combine it with window function $W(m_1, m_2; z)$ and redshift distribution $p(z)$ in Eq.~\eqref{eq:probability} to calculate a theoretical redshifted mass distribution. Then we can define an error function $E(\mathbf{n})$ by comparing theoretical distribution with the observed redshifted mass distribution at a selected set of PBH binary mass pairs as follows (see \cite{ciampiconi2023survey} for more details),
\begin{align}
	E(\mathbf{n}) \equiv \sqrt{\frac{\sum_{1 \leq i \leq j, \, j = 1 }^{j = N} [P_T(m_i^z, m_j^z) - P_O(m_i^z, m_j^z)]^2}{(1+N)N/2}}~,
\end{align}
where $P_{T}(m_1^z, m_2^z)$ and $P_{O}(m_1^z, m_2^z)$ are theoretical and observed redshifted mass distribution, respectively, and $N$ is the number of selected redshifted mass points for $m_1^z$ and $m_2^z$. 

The error function $E(\mathbf{n})$ only depends on the assumed PBH mass distribution vector $\mathbf{n}$, so any further derivation of $\mathbf{n}$ from the physical PBH mass function $n_p(m)$ would enhance the difference between $P_T(m_1^z, m_2^z)$ and $P_O(m_1^z, m_2^z)$ and increase the value of error function. It indicates that, in order to minimize the difference between assumed PBH mass distribution vector and the physical PBH mass function, minimizing the error function $E(\mathbf{n})$ could be an indicator to make $\mathbf{n}$ approaching the physical PBH mass function. To minimize the error function $E(\mathbf{n})$, we use the following iteration equation to update $\mathbf{n}$,
\begin{align}
	n_\mathrm{k+1}(m_i) = n_\mathrm{k}(m_i) - \gamma \frac{\partial E(\mathbf{n_\mathrm{k}})}{\partial n_\mathrm{k}(m_i)}~,
\end{align}
where $\mathbf{n}_\mathrm{k}$ is the $k \, $th iterated PBH mass vector and $n_\mathrm{k}(m_i)$ is its $i \,$th component. $\gamma$ is so-called learning rate in the gradient descent method, which describes the length of updated step in each iteration. With the update of $\mathbf{n}$, error function $E(\mathbf{n})$ is approaching its minimal value during iterations, the gradient of error function $\nabla E(\mathbf{n})$ is approaching zero vector, this causes the updated PBH mass distribution vector $\mathbf{n}_\mathrm{k+1} \approx \mathbf{n}_\mathrm{k}$. In this case, the iterated $\mathbf{n}$ can be viewed as a discretized approximation of the physical PBH mass function $n_p(m)$.
\begin{figure}[htbp] \centering
	\includegraphics[width=8cm]{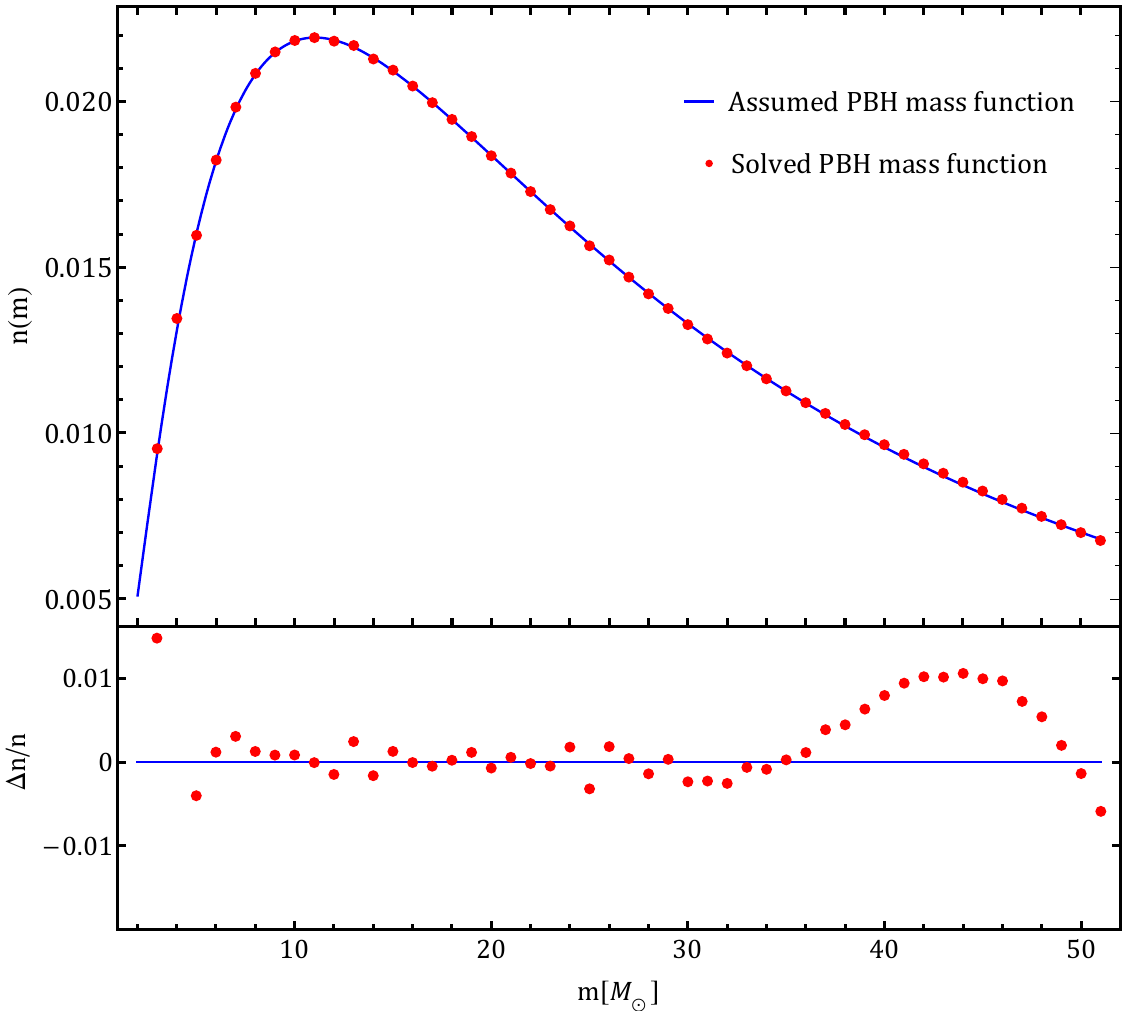}
	\caption{\label{fig:gradient_descent}
		\textit{Upper panel}: The data points are solved PBH mass distribution vector via the gradient descent method. The solid curve is the assumed log-normal PBH mass function with $m_c = 30 \, M_\odot$ and $\sigma = 1$.  \textit{Lower panel}: The relative error between solved PBH mass distribution vector and assumed PBH mass function. The background cosmology is a flat $\Lambda$CDM model with Planck 2018 parameters \cite{Planck:2018vyg}.
	}
\end{figure}

In Fig.~\ref{fig:gradient_descent}, we use the gradient descent method to inversely solve PBH mass function from a redshifted mass distribution of PBH binaries. This redshifted mass distribution is calculated under an assumed log-normal PBH mass function with $m_c = 30 \, M_\odot$ and $\sigma = 1$, which is shown as the solid curve, and the background cosmology is assumed as the flat $\Lambda$CDM model with Planck 2018 parameters \cite{Planck:2018vyg}. The data points are solved discrete PBH mass distribution vector, where we discretize PBH mass function into $50$ data points ranging from $1 \, M_\odot$ to $50 \, M_\odot$. In this iteration process, we obtain the minimal value in error function after $\mathcal{O}(50)$ times iterations, and we can find the solved PBH mass distribution vector can fit assumed PBH mass function to precision $\mathcal{O}(10^{-2})$. However, the relative error at small mass and large mass is relatively large compared with that at medium mass range. This is because the small mass tail and large mass tail only occupy a small part in PBH mass function and provide a small contribution in derived error function. This causes the constraints on small mass and large mass range is relatively weak via derivating the error function in the gradient descent method.

\section{Hubble parameter--dependent PBH mass function}\label{sec:Hubble_depend}

In extracting PBH mass function via inverse process from observed redshifted PBH mass function, the redshift distribution of detected PBH binaries $p(z)$ plays an important role in determining obtained PBH mass function. However, the extracted observables from PBH binaries in the GW channel are their redshifted mass $m_z$ and luminosity distance $d_L$, which cannot produce a redshift distribution of detected PBH binaries, due to lack of redshift information of PBH binaries. 

Although the redshift is degenerate with intrinsic mass in the observables of PBH binaries \cite{Schutz:1986gp, Chen:2017xbi}, we can still break this degeneracy by fixing some terms in the observables. As we can find that the luminosity distance $d_L(z)$ depends on the redshift of PBH binary and the cosmological model, if we fix an assumed cosmological model and its cosmological parameters, the redshift can be solved under such an assumption. Then, a model-dependent redshift distribution $p(z)$ can be constructed from the observables. The redshift and cosmological model dependence of luminosity distance in a flat cosmology \cite{Hogg:1999ad} is 
\begin{align}\label{eq:lumin_z_H0}
	d_L(z) =  (1+z) \int_0^z \frac{c}{H(z')} dz' = \frac{1+z}{H_0} \int_0^z \frac{c}{E_z(z')} dz'~,
\end{align}
where $H(z)$ is the Hubble parameter at different redshifts, $H_0$ is the present Hubble parameter, $E_z(z) = H(z)/H_0$ describes the evolution of Hubble parameter with respect to redshift. For any given cosmology with corresponding parameters, $E_z(z)$ would be fixed, then luminosity distance become a function of redshift and Hubble parameter $d_L \equiv d_L(z, H_0)$, we can also generalize this Hubble parameter dependence as $d_L \equiv d_L(z, H)$, where $H$ is the Hubble parameter at redshift in the range of $[0, z]$. In following discussion, we mainly focus on the dependence of present Hubble parameter $H_0$. 

Then redshift becomes a function of the luminosity distance and Hubble parameter $z(d_L, H_0)$, and can be determined from the luminosity distance by fixing a Hubble parameter. From a number of detected PBH binaries, we have a set of luminosity distances $\{d_L(z_i, H_0)| 1 \leq i \leq N\}$, where $N$ is the number of detected PBH binaries. For any given Hubble parameter $H_0$, a set of redshift can be obtained as $\{z_i = z(d_L^{\, i}, H_0)|1 \leq i \leq N\}$, which constructs a Hubble parameter-dependent redshift distribution $p(z; H_0)$. The dependence of redshift distribution on Hubble parameter behaves as follows,
\begin{align}\label{eq:redshift_dis_Hubble}
	p(z; \widetilde{H}_0) = \frac{1}{N_\mathrm{obs}}\frac{dN}{d\tilde{z}} = \frac{1}{N_\mathrm{obs}} \frac{dN}{dz} \frac{dz}{d\tilde{z}} = p(z; H_0) \frac{dz}{d\tilde{z}}~,
\end{align}
where $\widetilde{H}_0$ and $\tilde{z}$ are assumed Hubble parameter and its corresponding derived redshift, respectively. The differential relation of redshift $dz/d\tilde{z}$ can be found by derivating the relation $d_L(z, H_0) = d_L(\tilde{z}, \widetilde{H}_0)$, which gives,
\begin{align}
	\frac{dz}{d\tilde{z}} = \frac{\widetilde{D}_c + \widetilde{D}_H (1+\tilde{z})/E_z(\tilde{z})}{D_c + D_H (1+z)/E_z(z)}~.
\end{align}
Here, $D_c = D_H \int_0^z dz'/E_z(z)$ is the line-of-sight comoving distance \cite{Hogg:1999ad} and $\widetilde{D}_c$ is the line-of-sight comoving distance with an assumed Hubble parameter $\widetilde{H}_0$. Then, we can obtain the relation between redshift distribution with assumed Hubble parameter $p(z; \widetilde{H}_0)$ and the physical redshift distribution $p(z; H_0)$ from Eq.~\eqref{eq:redshift_dis_Hubble}.

In GW observation on PBH binaries, the intrinsic PBH mass cannot be determined from their observable, redshifted mass $m_z$ and luminosity distance $d_L$. However, if we can know the physical cosmological model and  Hubble parameter, the redshift can be determined from the luminosity distance, and the intrinsic PBH mass can be pinned down from mass-redshift degeneracy as follows,
\begin{align}\label{eq:intrinsic_mass}
	m = \frac{m_z}{1+z(d_L; H_0)}~.
\end{align}
It shows that PBH mass inferred from derived redshift in luminosity distance also depends on Hubble parameter, which indicates the PBH mass function should also be Hubble parameter-dependent. This can be found as follows, 
\begin{align}\label{eq:probability_2}
		P_O(m_1^z, m_2^z) = \int_0^\infty \tilde{n}(\frac{m_1^z}{1+z}) \tilde{n}(\frac{m_2^z}{1+z}) W(\frac{m_1^z}{1+z}, \frac{m_2^z}{1+z}; z) \frac{p(z; \widetilde{H}_0)}{(1+z)^2} \, dz~,
\end{align}
where  $p(z; \widetilde{H}_0)$ is a Hubble parameter-dependent redshift distribution under an assumed Hubble parameter $\widetilde{H}_0$ and $\tilde{n}(m)$ is the unknown PBH mass function, then we apply the gradient descent method in Eq.~\eqref{eq:probability_2} to inversely solve PBH mass function from observed redshifted mass distribution of PBH binaries as discussed in Sec.~\ref{sec:inverse}, and solved PBH mass function $\tilde{n}(m)$ should also be a Hubble parameter-dependent PBH mass function $n(m; \widetilde{H}_0)$.
\begin{figure}[htbp] \centering
	\includegraphics[width=8cm]{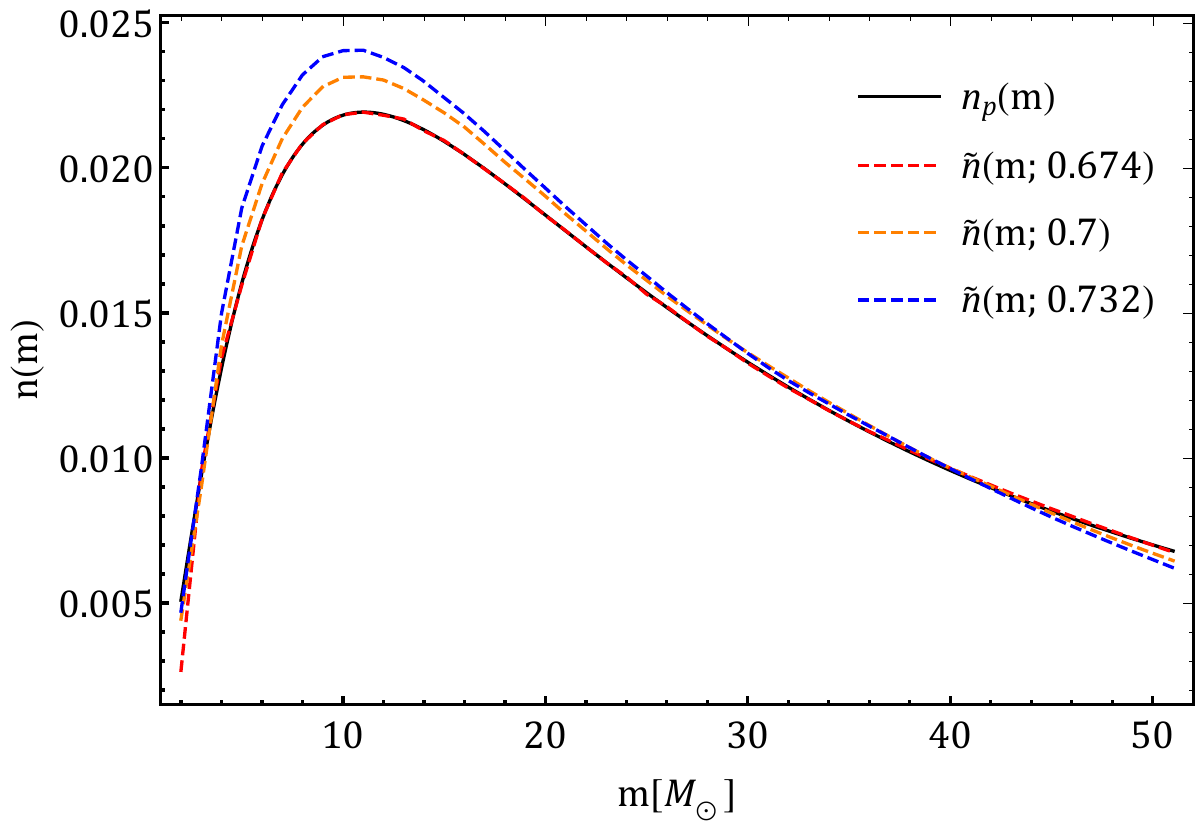}
	\caption{\label{fig:Hubble_dependent_mass_function}
		The Hubble parameter-dependent PBH mass functions and assumed physical PBH mass function. The solid curve is assumed physical PBH mass function with Hubble parameter $H_0 = 67.4 \, \mathrm{km} \, \mathrm{s}^{-1} \, \mathrm{Mpc}^{-1}$, this physical PBH mass function is a log-normal type with $m_c = 30 \, M_\odot$ and $\sigma = 1$. The dashed curves are inversely solved normalized PBH mass functions with different assumed Hubble parameters $h = H_0/(100 \, \mathrm{km} \, \mathrm{s}^{-1} \, \mathrm{Mpc}^{-1})$ as annotations. The background cosmology is a flat $\Lambda$CDM model with Planck 2018 parameters \cite{Planck:2018vyg}.
	}
\end{figure}

In Fig.~\ref{fig:Hubble_dependent_mass_function}, we can find the solved PBH mass function is Hubble parameter-dependent as we have discussed above. A larger assumed Hubble parameter would cause PBH mass distribution tend to smaller mass, this can be intuitively understood as a larger value of Hubble parameter gives a larger value of solved redshift in Eq.~\eqref{eq:lumin_z_H0}, and this larger redshift would cause a smaller value of intrinsic mass in Eq.~\eqref{eq:intrinsic_mass}. This figure shows that if we can have a constraint on the PBH mass function (we will discuss on it in Sec.~\ref{sec:merger_rate}), we can use it to measure the Hubble parameter. In such a measurement, measurement uncertainty is a key issue, which can be estimated based on the measurement precision of redshifted mass, intrinsic mass and luminosity distance in Eqs.~\eqref{eq:lumin_z_H0} and \eqref{eq:intrinsic_mass} as $\delta H_0 \sim \delta m_z - \delta m - \delta d_L$, where $\delta$ denotes the relative error. Currently, the measurement uncertainty of $m_z$ and $d_L$ is around $\mathcal{O}(10^{-1})$  in LIGO and Virgo (see TABLE VI of \cite{LIGOScientific:2020ibl}), if the measurement precision of intrinsic mass can achieve the same order as redshifted mass, this would causes the $\delta H_0 \sim \mathcal{O}(10^{-1})$. To further improve the measurement precision of the Hubble parameter, so that we can reveal the Hubble tension, the measurement uncertainty of Hubble parameter should achieve $\delta H_0 < 0.1$, this requires a large number of high SNR GW events to put a strong constraint on PBH mass function. With the development of GW detectors, especially space-based low frequency GW detectors, the measurement precision of extracted parameters can be further improved, which provides a higher measurement precision on Hubble parameter in this method and helps us understand the Hubble tension.

\section{Merger rate of PBH binaries as a probe of Hubble parameter}\label{sec:merger_rate}

The measurement of Hubble parameter requires a connection between different observables, where the scale of the universe and cosmic dynamics should be encoded \cite{Moresco:2022phi}, such as redshift-luminosity distance relation in standard candles \cite{Scolnic:2023mrv} and standard sirens \cite{Schutz:1986gp, Holz:2005df}, redshift-cosmic time relation in cosmic chronometers \cite{Stern:2009ep} and standard timers \cite{Cai:2021fgm, Ding:2022rpd}. In Eq.~\eqref{eq:probability_2}, such a relation is also connected, the redshift is encoded in observed $P_O(m_1^z, m_2^z)$, while cosmic dynamics is hidden in inferred $p(z; H_0)$, which requires an assumed cosmology and Hubble parameter to solve redshifts from the observed luminosity distances. In connecting the relation between $P_O(m_1^z, m_2^z)$ and $p(z; H_0)$, we need to know the physical PBH mass function. However, the PBH mass function is currently unknown, this causes the degeneracy between PBH mass function and redshift distribution, that is various combinations of PBH mass functions and redshift distributions can fit one observed $P_O(m_1^z, m_2^z)$, which causes the difficulty in determining Hubble parameter. Theoretically, we can calculate PBH mass function based on inflation models and PBH formation mechanisms, a lot of work has been done to derive PBH mass function \cite{Carr:1975qj, Lehmann:2018ejc, Suyama:2019npc, DeLuca:2020ioi, Bi:2022zdn}. Observationally, we have not proved the existence of the PBH, even though some GW events could be potential clues for PBHs \cite{LIGOScientific:2020aai, LIGOScientific:2020zkf, LIGOScientific:2020iuh}, and may help determine the PBH mass function \cite{Kimura:2021sqz, Wang:2022nml}. To determine the Hubble parameter from above relation, we  need to construct a relation bwtween an observable and PBH mass function and use this observable to constrain the PBH mass function. This relation must be independent on relation in Eq.~\eqref{eq:probability_2}, then it can be used to break the degeneracy between PBH mass function and redshift distribution and pin down the Hubble parameter.

The introduction of merger rate of PBH binaries $R$ changes the story, which provides another connection between an observable PBH merger rate and unknown PBH mass function. Combine this observable PBH merger rate with the relation in Eq.~\eqref{eq:probability_2}, the PBH mass function can be constrained and the Hubble parameter can be pinned down as a probe of Hubble parameter. It works as follows,
\begin{itemize}[itemsep=1pt,topsep=5pt,parsep=1pt]
	\item[1.] Give an assumed Hubble parameter $\widetilde{H}_0$ in a given cosmology.
	\item[2.] Use obtained $d_L$ from detected PBH binaries with $\widetilde{H}_0$ to construct $p(z; \widetilde{H}_0)$.
	\item[3.] Combine $p(z; \widetilde{H}_0)$ with observed $P_O(m_1^z, m_2^z)$ to solve $n(m; \widetilde{H}_0)$ via the gradient descent method.
	\item[4.] Calculate $R$ based on $n(m; \widetilde{H}_0)$ and compare it with observed PBH merger rate.
	\item[5.] Vary $\widetilde{H}_0$ until obtained $R$ matches the observed result, which pins down the correct Hubble parameter.
\end{itemize}

From the discussion in Sec.~(\ref{sec:redshifted_mass_dis}--\ref{sec:Hubble_depend}), we can use the gradient descent method to solve a Hubble parameter-dependent PBH mass function $n(m; \widetilde{H}_0)$ from an observed redshifted mass distribution $P_O(m_1^z, m_2^z)$. Then we can calculate PBH binary merger rate based on $n(m; \widetilde{H}_0)$ and compare it with observed PBH merger rate $R_O$ to find the best-fit Hubble parameter. To calculate the merger rate of PBH binaries, a number of works have been done, see \cite{Sasaki:2016jop, Ali-Haimoud:2017rtz, Raidal:2017mfl, Chen:2018czv, Liu:2018ess, Vaskonen:2019jpv, Raidal:2018bbj, Huang:2023klk} for details. We follow \cite{Chen:2018czv} and use the merger rate of PBH binaries $R$ with an extended mass function as,
\begin{align}\label{eq:merger_rate}
	R_{ij} = n_\mathrm{PBH} \min (n(m_i), n(m_j)) \Delta_m \frac{dP}{dt}~,
\end{align}
where, $n_\mathrm{PBH}$ is the present PBH number density, expressed as $n_\mathrm{PBH} = 3 H_0^2 \Omega_\mathrm{DM} f_\mathrm{PBH}/8 \pi G/\int m n(m) dm$ and $f_\mathrm{PBH}$ is the energy density fraction of PBHs in dark matter. $n(m_i)$ and $\Delta_m$ are the discretized PBH mass function and mass interval, respectively, which follow the normalization of probability as $\sum_{m_\mathrm{min}}^{m_\mathrm{max}} n(m_i) \Delta_m = 1$. $dP/dt$ is the probability distribution of merger time, see Appendix.~\ref{app:merger_rate} for detailed derivation.

\begin{figure}[htbp] \centering
	\includegraphics[width=8cm]{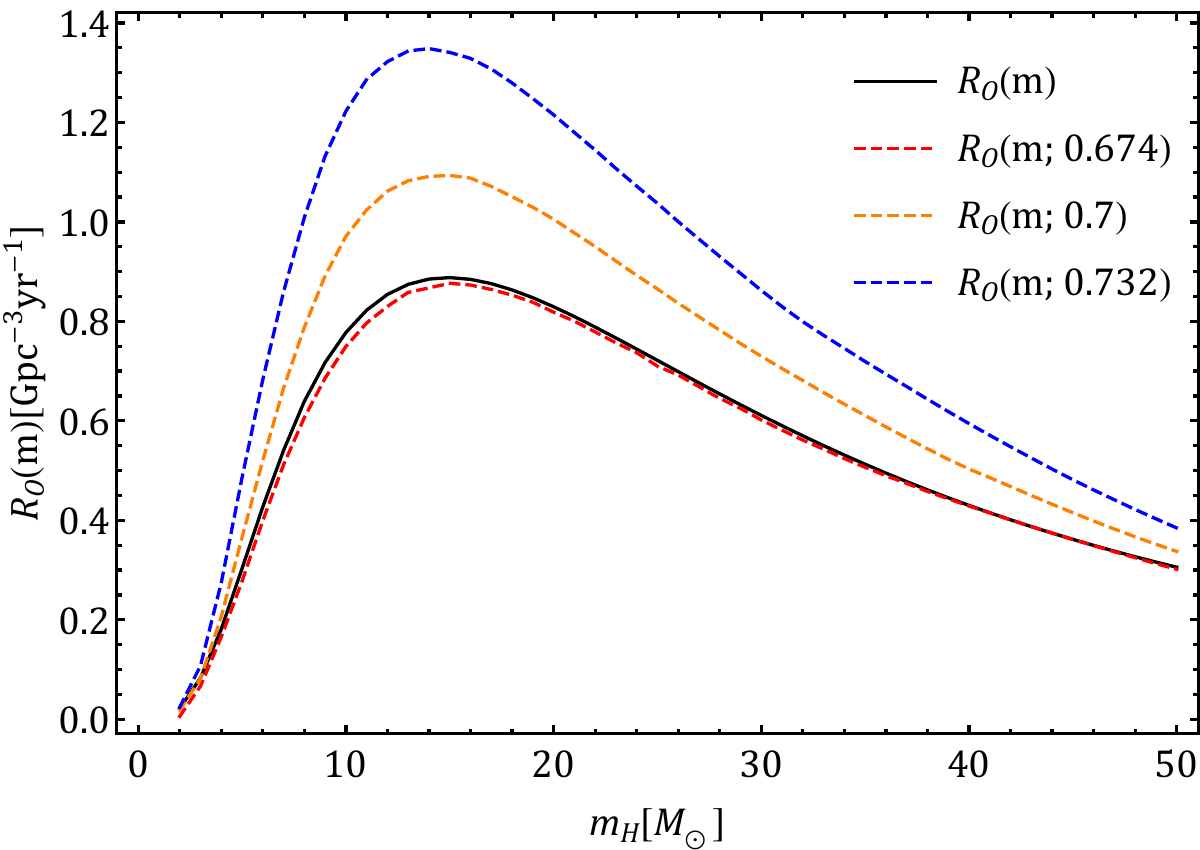}
	\includegraphics[width=8cm]{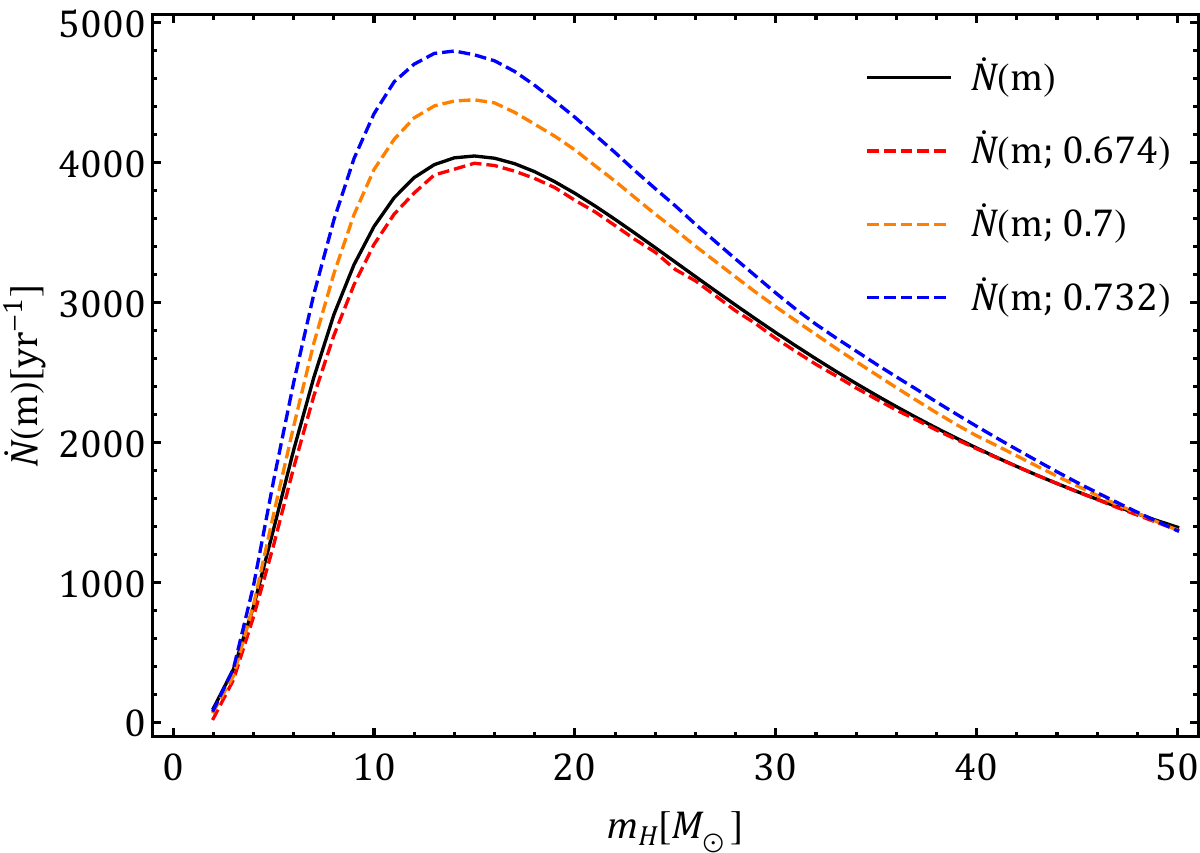}
	\caption{\label{fig:merger_rate}
		\textit{Left panel}: The comparison of observed merger rate distribution of PBH binaries at redshift $z = 20$ between assumed physical merger rate and calculated Hubble parameter-dependent merger rate. \textit{Right panel}: The comparison of merger rate population distribution of PBH binaries within redshift $z = 20$ and $z = 100$ between assumed physical merger rate population and calculated Hubble parameter-dependent merger rate population. The label of horizontal axis $m_\mathrm{H}$ denotes the mass of heavier PBH companion in PBH binaries. The solid curves are assumed physical merger rate distribution and merger rate population distribution, respectively. The assumed physical merger rate and population are calculated based on a log-normal PBH mass function with $m_c =  30 \, M_\odot$ and $\sigma = 1$. The dashed curves are calculated merger rate distribution and merger rate population distribution, respectively. The Hubble parameters denoted in dashed curves are $h = H_0/(100 \, \mathrm{km} \, \mathrm{s}^{-1} \, \mathrm{Mpc}^{-1})$. The background cosmology for both panels is a flat $\Lambda$CDM model with Planck 2018 parameters \cite{Planck:2018vyg} and the energy density fraction of PBHs in dark matter $f_\mathrm{PBH}$ is set as $f_\mathrm{PBH} = 0.001$. 
	}
\end{figure}

In Eq.~\eqref{eq:merger_rate}, the merger rate of PBH binaries depends on PBH mass function differently compared with the PBH mass function dependence in redshifted mass distribution, which indicates the merger rate of PBH binaries can provide an independent constraint on PBH mass function. As we have discussed above, the solved PBH mass function from inversing the redshifted mass distribution requires an assumed Hubble parameter, then we can use the solved Hubble parameter-dependent PBH mass function $n(m; \widetilde{H}_0)$ to calculate the merger rate distribution of PBH binaries theoretically, which should also be Hubble parameter-dependent $R(m_i, m_j; \widetilde{H}_0)$. Then we use the calculated observed PBH merger rate distribution at a given redshift $z$, where the observed merger rate is a redshift-diluted intrinsic merger rate, it follows $R_O = R/(1+z)$, to compare with the physical observed merger rate distribution. Only if we assumed the correct Hubble parameter, the theoretical result can match the observed result, and this comparison can help pin down the correct value of Hubble parameter, which is shown in the left panel of Fig.~\ref{fig:merger_rate}. Also, we can use the merger rate population of PBH binaries to constrain Hubble parameter-dependent PBH mass function, which can be calculated as follows,
\begin{align}
	\dot{N} = \int_{z_\mathrm{min}}^{z_\mathrm{max}}\frac{R}{1+z} \frac{dV_c}{dz} dz~,
\end{align}
where $\dot{N}$ denotes the merger rate population of PBH binaries between redshift $z_\mathrm{min}$ and $z_\mathrm{max}$, $dV_c/dz$ is the differential comoving volume as defined in Eq.~\eqref{eq:redshift_dis}. The comparison of merger rate population of PBH binaries between assumed physical population distribution and calculated population distribution is shown in the right panel of Fig.~\ref{fig:merger_rate}.

In Fig.~\ref{fig:merger_rate}, we calculate the distribution of merger rate and population of PBH binaries with respect to the mass of heavier PBH companion, which follow as $R(m_\mathrm{H}) = \sum_{i = 1}^{i = \mathrm{H}} R(m_i, m_\mathrm{H})$ and $\dot{N}(m_\mathrm{H}) = \sum_{i = 1}^{i = \mathrm{H}} \dot{N}(m_i, m_\mathrm{H})$. We can find that the calculated merger rate distribution and population distribution depend on the assumed value of Hubble parameter, where a larger value of Hubble parameter produces more merger rates of PBH binaries . This can be intuitively understood as a large value of Hubble parameter causes the PBH mass distribution tend to a smaller mass as we have discussed in Sec.~\ref{sec:Hubble_depend}, then a smaller intrinsic mass in PBH mass function would produce a larger merger rate in Eq.~\eqref{eq:merger_rate}. The difference in assumed value of Hubble parameter would produce distinct merger rate distributions, which can be used to pin down the correct Hubble parameter. As we can find in the both panels of Fig.~\ref{fig:merger_rate}, under an assumed flat $\Lambda$CDM model with $H_0 = 67.4 \, \mathrm{km} \, \mathrm{s}^{-1} \, \mathrm{Mpc}^{-1}$, we apply different values of Hubble parameter to inversely solve PBH mass function from the redshifted mass distribution of PBH binaries and calculate their merger rate distributions, only when Hubble parameter is approaching $ 67.4 \, \mathrm{km} \, \mathrm{s}^{-1} \, \mathrm{Mpc}^{-1}$, we can obtain the best-fit result. However, we still need to notice that this is a model-dependent method to measure the Hubble parameter, various background cosmologies would pin down different Hubble parameters. To achieve a better performance in probing Hubble parameter, a combination of this probe with other cosmic probes would provide a more precise measurement on Hubble parameter.

\section{Conclusion and Discussions}\label{sec:conclusion_discussion}

To summarize, we propose that the merger rate of PBH binaries can constrain the PBH mass function in the redshifted mass distribution of PBH binaries, and such a constraint on PBH mass function can help break the degeneracy between PBH mass function and redshift distribution and pin down the Hubble parameter. In GW observations on PBH binaries, a redshifted mass distribution of PBH binaries can be statistically constructed from redshifted mass of PBH binaries, and this redshifted mass distribution depends on the PBH mass function and the redshift distribution of detected PBH binaries, where the redshift distribution is not an observable in GW observations. To construct redshift distribution of detected PBH binaries, redshift need to be extracted from luminosity distances by assuming a value of the Hubble parameter in a given cosmology, which produces a Hubble parameter-dependent redshift distribution. In the connection among redshifted mass distribution, PBH mass function and the Hubble parameter-dependent redshift distribution, Hubble parameter cannot be extracted, due to the unknown PBH mass function. To extract Hubble parameter, we use the merger rate distribution of PBH binaries to provide an independent constraint on PBH mass function, which breaks this degeneracy between PBH mass function and redshift distribution and pins down the Hubble parameter.

In calculation, we focus on the redshifted mass distribution of PBH binaries at redshifts within $20 \leq z \leq 100$, where astrophysical BHs have not formed and  contributed, such a high redshift detection can be achieved in the next-generation GW detectors \cite{bender1998lisa, Harry:2006fi}. Then we can construct redshift distribution from a series of observed luminosity distances under an assumed Hubble parameter in a given cosmology (we use $\Lambda$CDM model in this work). To obtain PBH mass function from redshifted mass distribution, we apply the gradient descent method in the redshifted mass distribution to inversely solve the PBH mass function. Also, this solved PBH mass function depends on assumed Hubble parameter, due to the Hubble parameter dependence in redshift distribution. Based on solved Hubble parameter-dependent PBH mass function, we calculate the merger rate distribution of PBH binaries theoretically and compare it with observed result, where the compared observational merger rate distribution is calculated based on a log-normal PBH mass function with $m_c = 30 \, M_\odot$ and $\sigma = 1$ in a background of $\Lambda$CDM model with Planck 2018 parameters \cite{Planck:2018vyg}. To find the physical value of Hubble parameter, we vary the value of assumed Hubble parameter until the calculated merger rate distribution can match the observed result, where the best-fit Hubble parameter can be an approximation of its physical value.

In above discussion, we mainly focus on the contribution of intrinsic PBH mass function in the redshifted PBH mass function without considering the dynamical evolution of PBH mass function due to accretion effect \cite{DeLuca:2020bjf, DeLuca:2020qqa, DeLuca:2021pls} and the multiple mergers of PBH binaries \cite{Liu:2019rnx}. Although some dynamical effect can be ignored due to their negligible contribution \cite{Wu:2020drm, Liu:2022iuf}, a realistic Hubble parameter probe requires adding an extra contribution term in the physical PBH mass function in Eq.~\eqref{eq:probability_1}, see detailed discussion in Appendix.~\ref{app:modification}. Also, the measurement precision of Hubble parameter is a key issue in this method, its measurement uncertainty depends on the measurement uncertainty of redshifted mass, luminosity distance and PBH merger rate, that causes the precision of this method is not comparable with current methods, like standard candles and CMB power spectrum, as we have discussed in Sec.~\ref{sec:Hubble_depend}. For an applicable Hubble parameter measurement in this probe, it requires a measurement uncertainty for extracted parameters from GWs of PBH binaries, such as redshifted mass and luminosity distance, should achieve around $\mathcal{O}(10^{-2})$ in next-generation GW detectors.

To measure Hubble parameter, a number of methods could be applied in PBH binaries, such as gravitational-wave lensing \cite{Hannuksela:2020xor,  Wu:2022vrq, Huang:2023prq}, black hole population \cite{Chen:2022fda, LIGOScientific:2021aug} and statistic mass distribution in compact binaries \cite{Fung:2023yyq}, with these new methods and future development of GW observations, PBH binaries would bring us a high redshift measurement on Hubble parameter to reveal Hubble tension. In addition, PBH binaries can also provide a window in exploring the fundamental physics, e.g., the nature of dark matter \cite{Speeney:2022ryg, Pilipenko:2022emp, Akil:2023kym}, Hawking radiation \cite{Cai:2021zxo, Li:2020awx}, cosmic evolution \cite{Nakama:2020vtw}, etc. Therefore, pursuing these potential studies in PBH binaries could shed light on the nature of the universe and shows the great value of PBH binaries as a cosmic probe.

\section*{Acknowledgements}

I thank Masahide Yamaguchi, Otto A. Hannuksela, Xiao-Dong Li, Chao Chen, Leo W.H. Fung for useful discussions. I would like to thank Dr.~Xingwei Tang for nice discussions on the gradient descent method. This work is supported by the Institute for Basic Science under the project code, IBS-R018-D3.

\begin{appendix}

\section{The probability density distribution of redshifted PBH mass}\label{app:PBH_dis}

In order to calculate the probability density distribution on redshifted mass $P(m_1^z, m_2^z)$, we first express the cumulative distribution of redshifted PBH mass $C(m_1^z, m_2^z)$, which is a cumulative probability distribution on redshifted mass of observed PBH binaries that is defined as follows,
\begin{align}
	C(m_1^z, m_2^z) = \frac{N_\mathrm{obs}(m_a < m_1^z, m_b < m_2^z)}{N_\mathrm{tot}} = \int_0^{m_1^z} \int_0^{m_2^z} P(m_a, m_b) dm_a dm_b~,
\end{align}
where $N_\mathrm{obs}(m_a < m_1^z, m_b < m_2^z)$ is the number of observed PBH binaries with the redshifted mass of their components smaller than $m_1^z$ and $m_2^z$.

To express $C(m_1^z, m_2^z)$ in the form of PBH mass function, we first quantify the probability of PBH binaries with intrinsic mass $m_1$ and $m_2$ among all PBH binaries, we can start from calculating the number density of PBH binaries with intrinsic mass $(m_1, m_2)$, following Eq.~(2.18) in \cite{Raidal:2018bbj}
\begin{align}
	dn_b(m_1, m_2) = \frac{1}{2} e^{-\bar{N}(r)} n_\mathrm{PBH} n(m_1) dm_1 n_\mathrm{PBH} n(m_2) dm_2 dV~,
\end{align}
where $\bar{N}(r)$ is the expected number of PBH in a spherical volume with radius $r$, and the total number density of PBH binaries $n_b$ is
\begin{align}
	n_b = \int dn_b(m_1, m_2) = \frac{1}{2} e^{-\bar{N}(r)} n_\mathrm{PBH}^2 dV \int_0^\infty n(m_1) dm_1 \int_0^\infty n(m_2) dm_2~.
\end{align}
Then, we have the probability of PBH binaries with intrinsic mass $m_1$ and $m_2$ among all PBH binaries,
\begin{align}
	\frac{dn_b(m_1, m_2)}{n_b} = n(m_1) n(m_2) dm_1 dm_2~.
\end{align}
However, not all the PBH binaries with intrinsic mass $m_1$ and $m_2$ can be detected due to the limited sensitivity of GW detectors, it depends on their orbital parameters and corresponding redshift, so we use detectable window function $W(m_1, m_2; z)$ to give the detection probability of the observed PBH binaries among all existing PBH binaries with intrinsic mass $m_1$ and $m_2$ at redshift $z$ and these detected PBH binaries contribute the amount of PBH binaries with redshifted mass $m_1 (1+z)$ and $m_2 (1+z)$. The detected PBH binaries come from different redshifts, so their redshift have a distribution $p(z)$. Then we can have the redshifted mass contribution in $C(m_1^z, m_2^z)$ from PBH binaries with intrinsic mass $m_1$ and $m_2$ as
\begin{align}
	\frac{dC(m_1^z, m_2^z)}{dm_1 dm_2 dz} = n(m_1) n(m_2) W(m_1, m_2; z) p(z)~.
\end{align}
To evaluate $C(m_1^z, m_2^z)$, we first calculate its contribution within redshift range $(z, z+dz)$. For a given redshift $z$, the counted PBH binaries should satisfy that their intrinsic mass are in the range of $(0, m_1^z/1+z)$ and $(0, m_2^z/1+z)$, then we integrate the redshifted mass contribution over this intrinsic mass range, which can be expressed as follows,
\begin{align}
	\frac{dC(m_1^z, m_2^z)}{dz} = \int_0^{\frac{m_1^z}{1+z}} \int_0^{\frac{m_2^z}{1+z}} n(m_1) n(m_2) W(m_1, m_2; z) p(z) dm_1 dm_2~.
\end{align}
Then we accumulate all the counted PBH binaries from different redshifts to obtain $C(m_1^z, m_2^z)$ as,
\begin{align}\label{appeq:cumulative}
	C(m_1^z, m_2^z) = \int_0^\infty \int_0^{\frac{m_1^z}{1+z}} \int_0^{\frac{m_2^z}{1+z}} n(m_1) n(m_2) W(m_1, m_2; z) p(z) dm_1 dm_2 dz~.
\end{align}
Here we formally integrate redshift in the range of $(0, \infty)$. However, in practical, the bound of integration should be in a physical range like $(m_\mathrm{min}, m_1^z/1+z)$ which is determined by PBH mass distribution, and $(z_\mathrm{min}, z_\mathrm{max})$ which is determined by the GW detector. After obtaining $C(m_1^z, m_2^z)$, $P(m_1^z, m_2^z)$ can be calculated as $P(m_1^z, m_2^z) = \partial^2 C(m_1^z, m_2^z)/\partial m_1^z \partial m_2^z$, we first partial $C(m_1^z, m_2^z)$ with respect to $m_2^z$ as follows,
\begin{align}\label{appeq:cumu_1}\nonumber
	\frac{\partial C(m_1^z, m_2^z)}{\partial m_2^z} &= \lim_{\Delta m_2^z \to 0} \frac{1}{\Delta m_2^z} \int_0^\infty p(z) dz \int_0^{\frac{m_1^z}{1+z}} n(m_1) dm_1 \\\nonumber
	&\times \left(\int_0^{\frac{m_2^z+\Delta m_2^z}{1+z}} n(m_2) W(m_1, m_2; z)dm_2 - \int_0^{\frac{m_2^z}{1+z}} n(m_2) W(m_1, m_2; z)dm_2 \right)\\\nonumber
	&=\lim_{\Delta m_2^z \to 0} \frac{1}{\Delta m_2^z} \int_0^\infty p(z) dz \int_0^{\frac{m_1^z}{1+z}} n(m_1) dm_1 \int_{\frac{m_2^z}{1+z}}^{\frac{m_2^z+\Delta m_2^z}{1+z}} n(m_2) W(m_1, m_2; z)dm_2\\\nonumber
	&=\int_0^\infty p(z) dz \int_0^{\frac{m_1^z}{1+z}} n(m_1) dm_1 \times \lim_{\Delta m_2^z \to 0} \frac{1}{\Delta m_2^z} \int_{\frac{m_2^z}{1+z}}^{\frac{m_2^z+\Delta m_2^z}{1+z}} n(m_2) W(m_1, m_2; z)dm_2\\
	&=\int_0^\infty \int_0^{\frac{m_1^z}{1+z}} n(m_1) n(\frac{m_2^z}{1+z}) W(m_1, \frac{m_2^z}{1+z}; z) \frac{p(z)}{1+z} dm_1 dz~.
\end{align}
Then we partial Eq.~\eqref{appeq:cumu_1} with respect with $m_1^z$ as follows,
\begin{align}\label{appeq:cumu_2}\nonumber
	\frac{\partial}{\partial m_1^z} \left(\frac{\partial C(m_1^z, m_2^z)}{\partial m_2^z}\right) &= \lim_{\Delta m_1^z \to 0} \frac{1}{\Delta m_1^z} \int_0^\infty n(\frac{m_2^z}{1+z}) \frac{p(z)}{1+z} dz\\\nonumber
	&\times \left(\int_0^{\frac{m_1^z+\Delta m_1^z}{1+z}} n(m_1) W(m_1, \frac{m_2^z}{1+z}; z) dm_1 - \int_0^{\frac{m_1^z}{1+z}} n(m_1) W(m_1, \frac{m_2^z}{1+z}; z) dm_1 \right)\\\nonumber
	&=\lim_{\Delta m_1^z \to 0} \frac{1}{\Delta m_1^z} \int_0^\infty n(\frac{m_2^z}{1+z}) \frac{p(z)}{1+z} dz \int_{\frac{m_1^z}{1+z}}^{\frac{m_1^z+\Delta m_1^z}{1+z}} n(m_1) W(m_1, \frac{m_2^z}{1+z}; z) dm_1\\\nonumber
	&=\int_0^\infty n(\frac{m_2^z}{1+z}) \frac{p(z)}{1+z} dz \times \lim_{\Delta m_1^z \to 0} \frac{1}{\Delta m_1^z} \int_{\frac{m_1^z}{1+z}}^{\frac{m_1^z+\Delta m_1^z}{1+z}} n(m_1) W(m_1, \frac{m_2^z}{1+z}; z) dm_1\\
	&= \int_0^\infty n(\frac{m_1^z}{1+z}) n(\frac{m_2^z}{1+z}) W(\frac{m_1^z}{1+z}, \frac{m_2^z}{1+z}; z) \frac{p(z)}{(1+z)^2} \, dz~.
\end{align}
Then we obtain the expression of $P(m_1^z, m_2^z)$ as follows,
\begin{align}
	P(m_1^z, m_2^z) &= \frac{\partial^2 C(m_1^z, m_2^z)}{\partial m_1^z \, \partial m_2^z}\nonumber\\ 
	&= \int_0^\infty n(\frac{m_1^z}{1+z}) n(\frac{m_2^z}{1+z}) W(\frac{m_1^z}{1+z}, \frac{m_2^z}{1+z}; z) \frac{p(z)}{(1+z)^2} \, dz~.
\end{align}

\section{Late time modification on observed redshifted mass distribution of PBH binaries}\label{app:modification}

During the redshift evolution of PBH binaries, matter accretion from surrounding environment of PBH binaries would effectively increase the mass of PBHs, we follow \cite{DeLuca:2020fpg, Shapiro:1983du} and consider the accretion effect as
\begin{align}\label{eq:accretion}
	\frac{dm}{dt} = 4 \pi \lambda \rho_\mathrm{m} \frac{G^2 m^2}{v_\mathrm{eff}^3}~,
\end{align}
where $\rho_\mathrm{m}$ are the surrounding matter density of PBHs, $v_\mathrm{eff}$ is the PBH effective velocity, and accretion parameter $\lambda$ depends on surrounding matter density, the Hubble expansion rate, etc. Then we can find the redshift dependence of PBH mass by solving Eq.~\eqref{eq:accretion} as follows,
\begin{align}\label{eq:mz_mi}
	\int_{m_\mathrm{i}}^{m_z} \frac{dm}{m^2} = \frac{1}{m_\mathrm{i}} - \frac{1}{m_z} = \int_{z_\mathrm{i}}^{z} 4 \pi \lambda(z) \frac{G^2 \rho_\mathrm{m}}{v_\mathrm{eff}^3} \frac{dt}{dz} dz = g(z)-g(z_\mathrm{i})~,
\end{align}
where $dt/dz$ relation can be found under assuming cosmological model, and $g(z)$ is the integration of $4 \pi \lambda(z) \frac{G^2 \rho_\mathrm{m}}{v_\mathrm{eff}^3} \frac{dt}{dz}$. We also define $G(z,z_\mathrm{i}) \equiv g(z) - g(z_\mathrm{i})$ and use it in following discussion. Due to the PBH mass increment, PBH mass function at redshift $z$ can be expressed as follows,
\begin{align}
	n_z(m_z) = n_\mathrm{i}(m_\mathrm{i}) \frac{dm_\mathrm{i}}{dm_z} = n_\mathrm{i}(\frac{m_z}{1+G(z,z_\mathrm{i})m_z}) \frac{1}{(1+G(z, z_\mathrm{i})m_z)^2}~,
\end{align}
where $n_\mathrm{i}$ and $n_z$ are initial PBH mass function and PBH mass function at redshift $z$, respectively. $m_\mathrm{i}$ denotes the formation mass of PBHs and $m_z$ denotes the PBH mass at redshift $z$ after matter accretion effect. The differential relation $dm_\mathrm{i}/dm_z$ can be found from Eq.~\eqref{eq:mz_mi}. Then, the redshifted mass distribution of observed PBH binaries $P(m_1^z, m_2^z)$ is modified by this redshift-dependence PBH mass function $n_z(m_z)$.

Additionally, PBH binaries would also be disrupted by matter inhomogeneities as discussed in \cite{Raidal:2018bbj} or form by direct capture and three-body interaction inside the clusters \cite{Franciolini:2022ewd}. To quantify PBH binary disruption effect, we use suppression factor $S$ to calculate observed PBH binaries as follows,
\begin{align}
	\frac{dC_O(m_1^z, m_2^z)}{dm_1 dm_2 dz} = S \times \frac{dC(m_1^z, m_2^z)}{dm_1 dm_2 dz}~,
\end{align}
where $S \equiv S(m_1, m_2; f_\mathrm{PBH}, z)$ is a function of PBH intrinsic mass, and also depends on $f_\mathrm{PBH}$ and their corresponding redshift. A detailed expression for suppression factor in merger rate case can be found in Eq.~(2.37) of \cite{Raidal:2018bbj}. For our case, the expression of suppression factor of disrupted PBHs requires a further study in future. For late time PBH binaries formation by direct capture and three-body interaction inside the clusters, compare with early universe PBH binary formation, their number density can be negligible for $f_\mathrm{PBH} < 10^{-2}$, see Fig.~(4) of \cite{Franciolini:2022ewd}. 

Then, we can combine the matter accretion effect and disrupted effect to express Eq.~\eqref{eq:probability} as follows,
\begin{align}\nonumber
	P(m_1^z, m_2^z) =& \int_0^\infty n_z(\frac{m_1^z}{1+z}) n_z(\frac{m_2^z}{1+z}) S(\frac{m_1^z}{1+z}, \frac{m_2^z}{1+z}; f_\mathrm{PBH}, z) W(\frac{m_1^z}{1+z}, \frac{m_2^z}{1+z}; z) \frac{p(z)}{(1+z)^2} \, dz \nonumber\\
	=& \int_0^\infty n_\mathrm{i}(\frac{m_1^z}{1+z + G(z,z_\mathrm{i}) m_1^z}) n_\mathrm{i}(\frac{m_2^z}{1+z + G(z,z_\mathrm{i}) m_2^z}) \nonumber\\
	&\times \left(\frac{1+z}{1+z + G(z,z_\mathrm{i}) m_1^z}\right)^2 \left(\frac{1+z}{1+z + G(z,z_\mathrm{i}) m_2^z}\right)^2 S(\frac{m_1^z}{1+z}, \frac{m_2^z}{1+z}; f_\mathrm{PBH}, z) \nonumber\\ &\times W(\frac{m_1^z}{1+z}, \frac{m_2^z}{1+z}; z) \frac{p(z)}{(1+z)^2} \, dz~.
\end{align}
Once we obtain the detailed expression of $G(z,z_\mathrm{i})$ and $S(m_1, m_2; f_\mathrm{PBH}, z)$ in future study, the primordial mass function of PBHs can be numerically solved via the gradient descent method.

\section{The formation and merger rate of PBH binaries}\label{app:merger_rate}

In this section, we give a brief review on the formation and the merger rate of PBH binaries, see \cite{Ali-Haimoud:2017rtz, Chen:2018czv} for more details.

The PBH binary forms when two neighboring BHs are close enough and decouple from the Hubble flow. Given the equation of motion of two-point masses $M$ at rest with an initial separation $x$, the proper separation $r$ along the axis of motion evolves as
\begin{align}\label{eom}
	\ddot{r} - (\dot{H} + H^2) r + \frac{2M}{r^2}\frac{r}{|r|} = 0~,
\end{align}
where dots represent the differentiation with respect to proper time. In order to describe the early-universe evolution, we use $s \equiv a/a_{\mathrm{eq}}$ the scale factor normalized to unity at matter-radiation equality to express the Hubble parameter as $H(s) = (8 \pi \rho_{\mathrm{eq}}/3)^{1/2} h(s)$, where $h$ is defined as $h(s) \equiv \sqrt{s^{-3} + s^{-4}}$ and $\rho_{\mathrm{eq}}$ is the matter density at equality. Then, we can rewrite Eq.~\eqref{eom} by introducing $\chi \equiv r/x$ as
\begin{align}
\chi'' + \frac{sh' + h}{s^2 h}(s \chi' - \chi) + \frac{1}{\lambda}\frac{1}{(sh)^2}\frac{1}{\chi^2}\frac{\chi}{|\chi|} = 0~,
\end{align}
where primes denote differentiation with respect to $s$. Here, the dimensionless parameter $\lambda$ is defined as $\lambda \equiv 4 \pi \rho_{\mathrm{eq}} x^3/3 M$. The initial condition is given in the condition that the two neighboring BHs follow the Hubble flow $\chi(s) = s$, the initial conditions are
\begin{align}
\chi(0) = 0~,~~~\chi'(0) = 1~.
\end{align}
Then, the numerical solution in \cite{Ali-Haimoud:2017rtz} shows that the binary effectively decouples from the Hubble flow at $s \approx \lambda/3$. The corresponding redshift is
\begin{align}\label{appen_eq:formation_redshift}
z = \frac{3(1 + z_{\mathrm{eq}})}{\lambda} - 1~.
\end{align}
In order to get the merger rate of PBH binaries with mass distribution $n(m)$ at cosmic time $t$, we follow \cite{Chen:2018czv} and start from discretizing the PBH mass function $n(m)$ to define a binned mass distribution $n(m_i)$ and mass interval $\Delta_m$, which follows
\begin{align}
	\sum_{m_\mathrm{min}}^{m_\mathrm{max}} n(m_i) \Delta_m = 1~.
\end{align}
Then, the average distance between two nearby PBHs is
\begin{align}\label{eq:mean_distance}
	\langle x_{ij} \rangle = (\bar{x}_i^{-3} + \bar{x}_j^{-3})^{-1/3} = \mu_{ij}^{1/3} \bar{x}_{ij}~,
\end{align}
where, $\mu_{ij}$ and $\bar{x}_{ij}$ are defined as
\begin{align}\label{appen_eq:mean_dist}
\mu_{ij} = \frac{2 (n(m_i)m_i + n(m_j) m_j)}{m_b (n(m_j) + n(m_i))}~,~~~\bar{x}_{ij}^3 = \frac{3}{8 \pi} \frac{m_b}{\rho_\mathrm{eq} f_b \Delta_m}~,
\end{align}
where, $f$ is the fraction of PBH energy density in matter, $f_b = f n_\mathrm{PBH} (n(m_i) m_i + n(m_j) m_j)/\rho_\mathrm{PBH}$ and $m_b = m_i + m_j$. $\rho_\mathrm{eq} = \Omega_\mathrm{M} \rho_\mathrm{crit} (1+z_\mathrm{eq})^3$ is the total energy density of matter at matter-radiation equality. After the formation of PBH binaries, the orbit of PBH binary shrinks due to the emission of GWs, which leads the merger of PBH binary, the coalescence time of the PBH binary is given in \cite{Peters:1964zz} by,
\begin{align}\label{coal_time}
	t = \frac{3}{85} \frac{a^4}{G^3 m_1 m_2 (m_1 + m_2)} j^7~,
\end{align}
From Eq.~\eqref{coal_time}, the merger rate of PBH binaries can be obtained from the initial semi-major axis $a$ distribution and the initial dimensionless angular momentum $j$ distribution (the dimensionless angular momentum can be represented in term of eccentricity of PBH binary $e$ as $j = \sqrt{1-e^2}$). The initial semi-major axis $a$ after the binary formation is numerically given in \cite{Ali-Haimoud:2017rtz},
\begin{align} \label{major_axis}
a \approx 0.1 \lambda x = \frac{0.1 \bar{x}_{ij}}{f_b \Delta_m}X^{4/3},
\end{align}
where, $X \equiv x^3/\bar{x}_{ij}^3$.
Therefore, the initial semi-major axis distribution is determined by the separation $x$ distribution, we follow \cite{Sasaki:2016jop, Ali-Haimoud:2017rtz, Chen:2018czv} that assuming PBHs follow a random distribution, the probability distribution of the separation $x$ is
\begin{align}
\frac{dP}{dX} = \mu_{ij}^{-1} e^{-X\frac{4\pi}{3}\bar{x}_{ij}^3n_{T}}~,
\end{align}
where $n_{T} \equiv n_{\mathrm{PBH}}(1+z_{\mathrm{eq}})^3\int_0^{\infty} n(m)dm$. Considering fixed $X$, the dimensionless angular momentum $j$ can be given by Eq.~\eqref{coal_time} and Eq.~\eqref{major_axis},
\begin{align}
j(t;X) = \left( \frac{3}{85} \frac{G^3 m_1 m_2 (m_1 + m_2) (f_b \Delta_m)^4}{(0.1 \bar{x}_{ij})^4 X^{16/3}} t \right)^{1/7}~.
\end{align}
The differential probability distribution of $(X, t)$ is given by
\begin{align} \label{xt_dis}
	\frac{d^2P}{dXdt} = \frac{dP}{dX}\left(  \frac{\partial j}{\partial t}\frac{dP}{dj} \bigg|_X \right)_{j(t;X)} = \frac{\mu_{ij}^{-1}}{7t}e^{-X\frac{4\pi}{3}\bar{x}_{ij}^3n_{T}}\mathcal{P}(j/j_X)~,
\end{align}
where $j_X = 0.5 f X/ f_b \Delta_m$, $\mathcal{P}(j/j_X) = (j/j_X)^2/(1 + j^2/j_X^2)^{3/2}$. Integrating Eq.~\eqref{xt_dis} gives the merger time probability distribution
\begin{align}
\frac{dP}{dt} = \frac{\mu_{ij}^{-1}}{7t}\int dX e^{-X\frac{4\pi}{3}\bar{x}_{ij}^3n_{T}}\mathcal{P}(j/j_X)~.
\end{align}
Then, the comoving merger rate $R_{ij}$ for PBH binaries at time $t$ is
\begin{align}
R_{ij}(t) =  n_{\mathrm{PBH}}\mathrm{min}(n(m_i),n(m_j))\Delta_m \frac{dP}{dt}~.
\end{align}

\end{appendix}

\end{document}